\documentclass[a4paper,11pt]{article}
\usepackage{pos}
\usepackage{xspace}

\newcommand{\lam}{\ensuremath{\Lambda}\xspace}
\newcommand{\alam}{\ensuremath{\bar{\Lambda}}\xspace}

\title{Hyperon polarization measurements in heavy-ion collisions}

\author*{Takafumi Niida}

\affiliation{Department of Physics, Institute of Pure and Applied Sciences, University of Tsukuba\\
  1-1-1 Tennoudai, Tsukuba, Ibaraki 305-8571, Japan}

\emailAdd{niida.takafumi.fw@u.tsukuba.ac.jp}

\abstract{
In non-central heavy-ion collisions, a large orbital angular momentum of the colliding system is produced, which is then partially transferred to the created medium, resulting in the particle polarization on average along the initial angular momentum, known as global polarization. It was predicted almost 20 years ago and the first observation of $\Lambda$ global polarization was made by the STAR Collaboration in 2017. Since then, a lot of progress have been made in the polarization measurements including global polarization of multistrange hyperons and the polarization along the beam direction induced by azimuthal anisotropic flow. In these proceedings, we present recent experimental progress on the hyperon polarization measurements in heavy-ion collisions.
}

\FullConference{25th International Spin Physics Symposium (SPIN 2023)\\
 24-29 September 2023\\
 Durham, NC, USA\\}

\begin{document}
\maketitle
\section{Introduction}
Heavy-ion collision experiments have been conducted at the Relativistic Heavy Ion Collider (RHIC) and the Large Hadron Collider (LHC) to study the properties of quark-gluon plasma as well as to understand the nature of the phase structure of quantum chromodynamics (QCD). The discovery of global polarization of \lam hyperons by the STAR Collaboration in 2017~\cite{STAR:2017ckg} has opened a new direction in the field of heavy-ion collision physics. Since then, the polarization measurements and theoretical study on spin dynamics become one of hot topics and a lot of progress have been made. In these proceedings, we present recent experimental results on hyperon polarization measurements in heavy-ion collisions, discuss their physics implications and open questions, and briefly mention the outlook on the polarization measurements in heavy-ion collisions.


\section{Hyperon polarization measurements}
Particle polarization can be measured by utilizing the nature of hyperon's weak decay where the parity is not conserved. The distribution of daughter particle in the rest frame of the hyperon depends on the hyperon polarization as shown below:
\begin{equation}
\frac{dN}{d\Omega^\ast} = \frac{1}{4\pi} (1+\alpha_H {\mathbf P}_H^\ast \cdot \hat{\mathbf p}_B^\ast),
\end{equation}
where ${\mathbf P}_H$ is the hyperon polarization vector and $\hat{{\mathbf p}}_B$ is the unit vector of the daughter baryon momentum where the asterisk denotes the rest frame of the hyperon. The factor $\alpha_H$ is the decay parameter of hyperon reflecting the sensitivity of the measurement: $\alpha_{\Lambda}= 0.732 \pm 0.014$ for $\Lambda$ and $\alpha_{\Xi}=-0.401\pm 0.010$ for $\Xi^{-}$~\cite{ParticleDataGroup:2022pth}. In case for multistrange hyperons such as $\Xi$ and $\Omega$, they decay in two steps, e.g., $\Xi^{-}\rightarrow\Lambda\pi^{-}$ and then $\Lambda\rightarrow p\pi^{-}$. If the parent particle is polarized, its polarization is transferred to the daughter particle depending on the type of decay. In case for weak decays of spin $1/2$ ($\Xi$) and $3/2$ ($\Omega$) hyperons, the daughter $\Lambda$ polarization can be expressed by the parent hyperon polarization with other decay parameter $\gamma$ as follows~\cite{Lee:1957qs,Becattini:2016gvu}:
\begin{eqnarray}
P_{\Lambda}^\ast &=& C_{\Xi\Lambda} P_{\Xi}^{\ast} = \frac{1}{3}(1+2\gamma_{\Xi}) P_{\Xi}^\ast = 0.944 P_{\Xi}^\ast, \\
P_{\Lambda}^\ast &=& C_{\Omega\Lambda} P_{\Omega}^{\ast} = \frac{1}{5}(1+4\gamma_{\Omega}) P_{\Omega}^\ast,
\end{eqnarray}
where $C_{\Xi\lam}$ and $C_{\Omega\lam}$ are polarization transfer factor. The decay parameter $\gamma_{\Omega}$ is unknown but is expected to be $\gamma_{\Omega}\approx\pm1$ with the ambiguity of the sign~\cite{Luk:1988as}, which leads to $C_{\Omega\Lambda}\approx1$ or $C_{\Omega\Lambda}\approx-0.6$. In other words, based on the global polarization picture, one can constrain the unmeasured $\gamma_{\Omega}$ by measuring global polarization of $\Omega$ hyperons.

\section{Polarization along the initial angular momentum}
In non-central heavy-ion collisions, the colliding system carries an initial orbital angular momentum of the order of $L\!\sim\!10^6\hbar$ which is partially transferred to  the created medium. As a consequence, produced particles are globally polarized on average along the initial angular momentum via spin-orbit coupling~\cite{Liang:2004ph,Voloshin:2004ha,Becattini:2007sr}; the phenomenon is referred to as ``global polarization". Assuming the local thermal equilibrium, the polarization can be expressed with the local thermal vorticity~\cite{Becattini:2007sr}. In the nonrelativistic limit, the polarization $\mathbf P$ can be written by:
\begin{equation}
{\mathbf P} \approx \frac{(s+1)}{3} \frac{({\mathbf \omega}+\mu {\mathbf B}/s)}{T},
\label{eq:Pol}
\end{equation}
where $s$ and $\mu$ are spin and magnetic moment of particle, $T$ is the temperature, and $\mathbf B$ is the magnetic field at the freeze-out. Based on Eq.~\eqref{eq:Pol}, one can expect that particle with different spin get polarized differently and there might be the difference between particle and antiparticle due to the opposite sign of $\mu$ if the magnetic field contribution is significant in the measured polarization.

Polarization along the initial angular momentum of the colliding system can be defined as~\cite{STAR:2007ccu}
\begin{equation}
P_H = \frac{8}{\pi \alpha_H} \frac{1}{A_0} \frac{\langle\sin(\Psi_1^{\rm obs}-\phi_B^\ast)\rangle}{\rm Res(\Psi_1)},
\end{equation}
where $A_0$ is an acceptance correction factor close to unity depending on multiplicity and particle momentum, $\Psi_1$ is azimuthal angle of the first-order event plane as a proxy of reaction plane, and $\phi_B^\ast$ is azimuthal angle of the daughter baryon in the hyperon rest frame. The Res($\Psi_1$) in the denominator represents the event plane resolution (unity for a perfect detector).

\subsection{Energy dependence}
\begin{figure}[htbp]
\begin{center}
\includegraphics[width=0.7\linewidth]{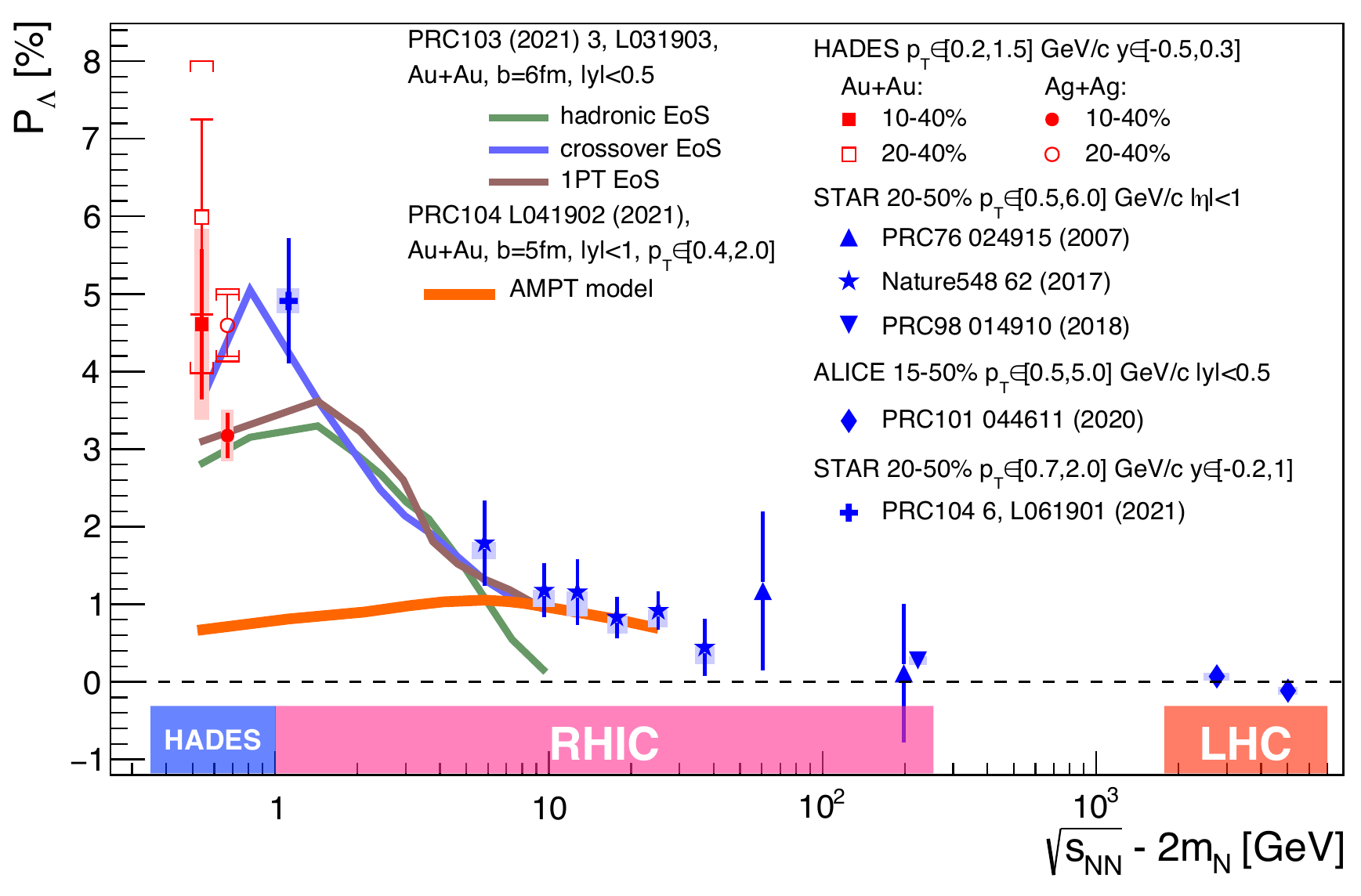}
\caption{$\Lambda$ global polarization as a function of the center-of-mass energy per nucleon pair, $\sqrt{s_{NN}}-2m_N$, where $m_N$ is the nucleon mass. Calculations from 3D-fluid-dynamics with different EoS and a multi-phase transport (AMPT) model are shown~\cite{Ivanov:2020udj}. The figure is taken from Ref.~\cite{HADES:2022enx}.}
\label{fig:PLam-rootS}
\end{center}
\end{figure}

Global polarization of \lam hyperons has been measured in a wide range of collision energy from a few GeV to a few TeV since the first observation by the STAR Collaboration in the beam energy scan phase-I (BES-I)~\cite{STAR:2017ckg,STAR:2018gyt,ALICE:2019onw,STAR:2021beb,HADES:2022enx,STAR:2023nvo}. Figure~\ref{fig:PLam-rootS} shows \lam global polarization as a function of the center-of-mass energy per nucleon pair $\sqrt{s_{NN}}$ for mid-central heavy-ion collisions, where the increasing trend towards lower energies can be seen. Such an energy dependence can be explained well by theoretical models such as hydrodynamic and transport models (see recent reviews, e.g., Refs.~\cite{Becattini:2020ngo,Becattini:2021wqt} and references therein).
The observed energy dependence can be understood by larger shear flow structure at midrapidity in the initial state due to baryon stopping in lower energies and less dilution effect of the system vorticity due to shorter system lifetime~\cite{Karpenko:2016jyx}.
Theoretical models predict that the global polarization peaks around a few GeV where the rise starts at the energy near the threshold of nucleon pair production ($2m_N$)~\cite{Deng:2020ygd,Ayala:2021xrn,Guo:2021udq}.
Recent results from STAR fixed-target data at 3 GeV~\cite{STAR:2021beb} as well as HADES results at 2.4 GeV and 2.55 GeV~\cite{HADES:2022enx} indicate that the global polarization seems to continue increasing at these energies, although the uncertainties are still large and to be improved in near future.

\subsection{Difference between particle and antiparticle}
\begin{figure}[htbp]
\begin{center}
\includegraphics[width=0.4\linewidth]{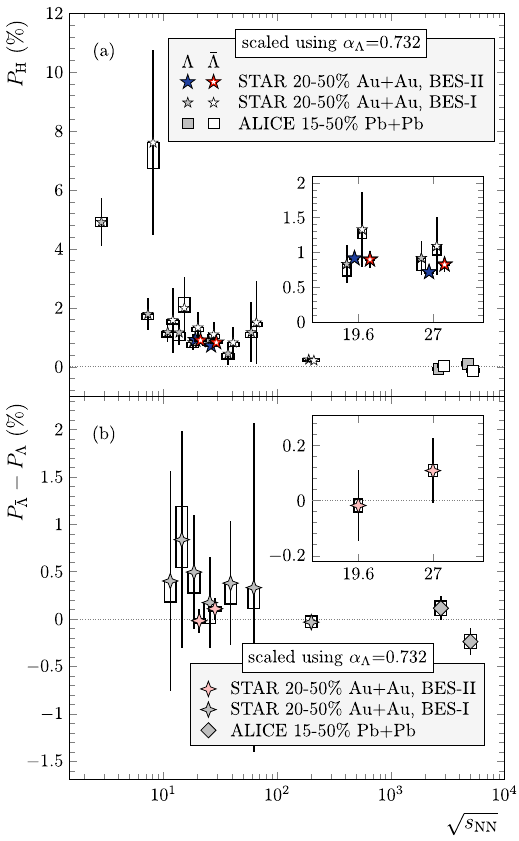}
\hspace{0.4cm}
\includegraphics[width=0.5\linewidth]{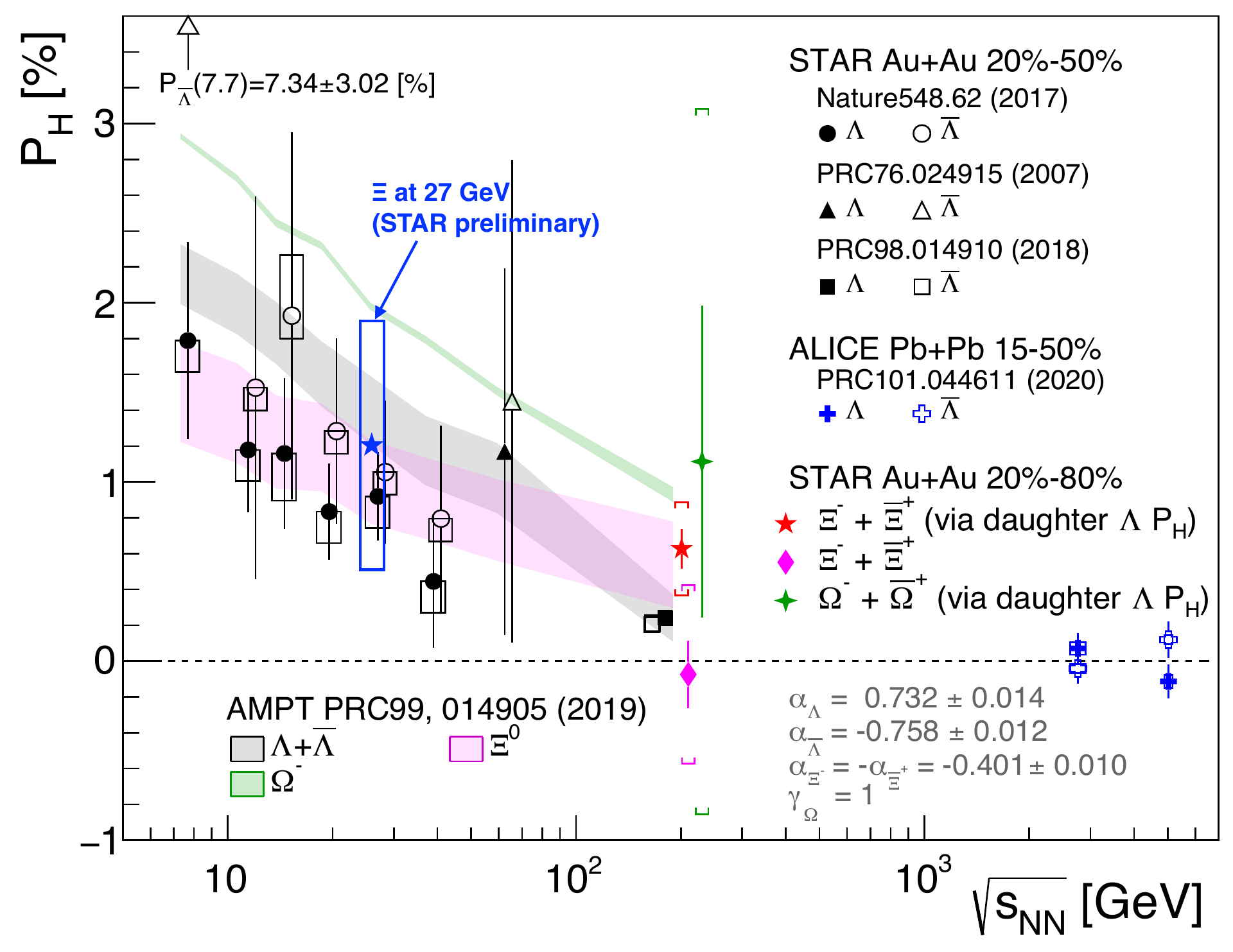}
\caption{(Left) Global polarization of \lam and \alam hyperons as a function of $\sqrt{s_{NN}}$ in the top panel and the polarization difference $P_{\bar{\Lambda}}-P_{\Lambda}$ in the bottom panel with new results at 19.6 GeV and 27 GeV from BES-II at RHIC. (Right) Global polarization of $\Xi$ and $\Omega$ hyperons at $\sqrt{s_{NN}}$ = 200 GeV as well as that of \lam (\alam). These figures are taken from Refs.~\cite{STAR:2023nvo,STAR:2020xbm}.}
\label{fig:DPH-rootS}
\end{center}
\end{figure}

In non-central heavy-ion collisions, the strong magnetic field, of the order of $|B|\sim10^{14}$~T, would be created in the initial state because electric charges in the spectator fragments move in the opposite direction nearly at the speed of light. The magnetic field points to the direction perpendicular to the reaction plane, which coincides with the initial angular momentum direction. Therefore the magnetic field may contribute to the global polarization, resulting in the splitting of the polarization between particles and antiparticles as indicated in Eq.~\eqref{eq:Pol}. Figure~\ref{fig:DPH-rootS} (left) shows \lam and \alam global polarization and their difference, $P_{\alam}-P_{\lam}$, as a function of $\sqrt{s_{NN}}$, with new results at 19.6 GeV and 27 GeV from STAR in BES-II. The uncertainties are greatly reduced compared to the results in BES-I because of large statistics and detector upgrades but there is no significant difference between \lam and \alam. Since the lifetime of the magnetic field is expected to be very short ($<0.5$~fm/$c$)~\cite{McLerran:2013hla}, one may not expect any significant contribution from the magnetic field to the global polarization. Nevertheless one can estimate the upper limit of the late-stage magnetic field as $|B|\approx T|P_{\alam}-P_{\lam}|/(2|\mu_{\Lambda}|)$~\cite{Becattini:2016gvu,Muller:2018ibh} and find it to be $|B|<10^{13}$~T from the BES-II results taking the temperature $T$ being 150~MeV. Note that the lifetime of the magnetic field is sensitive to the electric conductivity of the medium~\cite{McLerran:2013hla}.

Recently, the STAR Collaboration studied the global polarization in isobar $^{96}_{44}$Ru+$^{96}_{44}$Ru and $^{96}_{40}$Zr+$^{96}_{40}$Zr collisions at $\sqrt{s_{NN}}$ = 200 GeV~\cite{STAR:2021mii} where the initial magnetic field squared would be about 15\% larger in Ru+Ru because of its larger atomic number. Preliminary results in isobar collisions show no significant difference between \lam and \alam as well as the two isobar collisions~\cite{Guo-spin2023}.

\subsection{Multistrange hyperons}
In order to better understand the polarization mechanism, it is important and of great interest to study different particle species with different spin. Global polarization of multistrange hyperons, i.e., $\Xi^{-}(\bar\Xi^{+})$ and $\Omega^-(\bar\Omega^{+})$, was also studied by STAR in Au+Au collisions at $\sqrt{s_{NN}}$ = 200 GeV~\cite{STAR:2020xbm}. Global polarization of $\Xi$ hyperons averaged over the two independent methods is found to be $\langle P_{\Xi}\rangle = 0.47\pm0.10\pm0.23$ (\%), which confirms the global polarization picture based on the fluid vorticity. The preliminary result of $P_{\Xi}$ at 27 GeV~\cite{Alpatov:2020iev} seems to also follow the global trend of the energy dependence. As shown in Fig.~\ref{fig:DPH-rootS} (right), the data show a hint of hierarchy in the observed polarization, i.e., $P_{\Omega}>P_{\Xi}>P_{\Lambda}$, which can be interpreted as combinations of spin dependence as in Eq.~\eqref{eq:Pol} and the feed-down effect~\cite{Li:2021zwq}. The result on $\Omega$ polarization has still large uncertainty and will be improved with the coming new data in 2023+2025 RHIC runs which also helps us to shed light on the ambiguity on the sign of the decay parameter $\gamma_{\Omega}$.

\section{Polarization along the beam direction}
\begin{figure}[htbp]
\begin{center}
\includegraphics[width=1.01\linewidth]{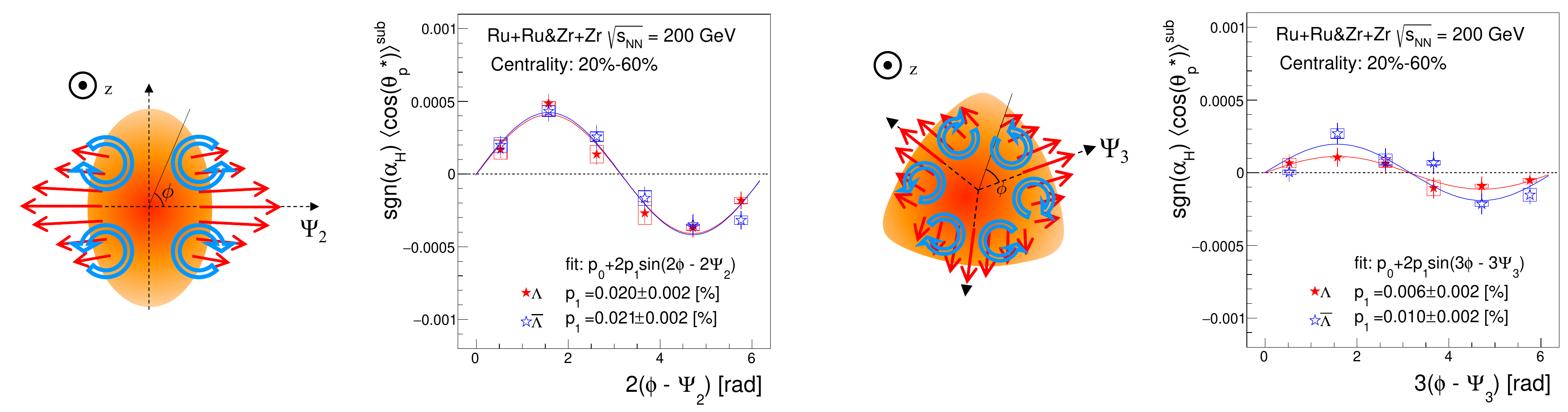}
\caption{Polarization of \lam and \alam hyperons along the beam direction, $P_z\propto \langle\cos\theta_p^\ast\rangle$, as a function of hyperons' azimuthal angle relative to the second- (left panel) and third-order (right panel) event planes in isobar collisions at $\sqrt{s_{NN}}$ = 200 GeV. Cartoons next to the plots depict the initial geometry viewed from the beam direction (z-direction) with flow velocity due to anisotropic flow (solid arrows) and the consequent vorticities (open arrows).
These figures are taken from Ref.~\cite{STAR:2023eck}.}
\label{fig:PzvsDPhi}
\end{center}
\end{figure}

It has been suggested that complex vortical structures such as a toroidal vortex could be created in addition to the global vorticity in heavy-ion collisions due to collective expansion~\cite{Pang:2016igs,Xia:2018tes}, jet-medium interaction~\cite{Betz:2007kg,Serenone:2021zef}, and in asymmetric collisions~\cite{Voloshin:2017kqp,Lisa:2021zkj}. Refs.~\cite{Voloshin:2017kqp,Becattini:2017gcx} predicted that polarization along the beam direction could be induced due to anisotropic flow. As shown in the left cartoon of Fig.~\ref{fig:PzvsDPhi}, expansion velocity in the transverse plane depends on the azimuthal angle and becomes stronger in in-plane direction (shorter axis of the ellipse) for the elliptic initial geometry, leading to the vorticity and therefore the polarization along the beam direction. Polarization along the beam direction can be defined as follows~\cite{STAR:2019erd}:
\begin{equation}
P_z = \frac{\langle\cos\theta_B^\ast\rangle}{\alpha_H \langle\cos^2\theta_B^\ast\rangle} \approx \frac{3\langle\cos\theta_B^\ast\rangle}{\alpha_H},
\end{equation}
where $\theta_B^\ast$ is the polar angle of daughter baryon in the hyperon rest frame relative to the beam direction. The factor $\langle\cos^2\theta_B^\ast\rangle$ accounts for the acceptance effect and is estimated in a data-driven way, usually close to $1/3$ as it should be for the perfect detector.

The STAR Collaboration observed such an anisotropic-flow-driven polarization with respect to the second-order event plane (plane for elliptic flow) in Au+Au collisions at $\sqrt{s_{NN}}$ = 200 GeV~\cite{STAR:2019erd}. It was also confirmed later by ALICE in Pb+Pb collisions at $\sqrt{s_{NN}}$ = 5.02 TeV~\cite{ALICE:2021pzu}. 
However, there has been discrepancy in the sign of the polarization between the data and many models, even among hydrodynamic models with different approaches. Also, a simple hydrodynamics-inspired blast-wave model is found to explain the data reasonably~\cite{Voloshin:2017kqp,STAR:2019erd}. The situation is referred to as the ``spin sign puzzle" in heavy-ion collisions (see Refs.~\cite{Becattini:2020ngo,Becattini:2021wqt} for review). Recent theoretical studies~\cite{Fu:2021pok,Becattini:2021iol} suggest that the inclusion of the shear-induced polarization (SIP) is needed to explain the data, although the result depends on the detailed implementation.

New results using the isobar data have been recently released by STAR, extending the measurement to higher harmonic flow~\cite{STAR:2023eck}. Figure~\ref{fig:PzvsDPhi} shows raw signals of polarization along the beam direction (before the correction for the event plane), $\langle\cos\theta_p^\ast\rangle \approx \alpha_H P_z/3$, as a function of hyperons' azimuthal angle relative to the second- (left panel) and third-order (right panel) event planes in isobar collisions at $\sqrt{s_{NN}}$ = 200 GeV. As seen for the second-order case, a similar sine pattern of the polarization was observed for the third-order case, indicating a triangular-flow-driven sextupole pattern of vorticity as depicted in the right cartoon of Fig.~\ref{fig:PzvsDPhi}.

\begin{figure}[htbp]
\begin{center}
\includegraphics[width=0.85\linewidth]{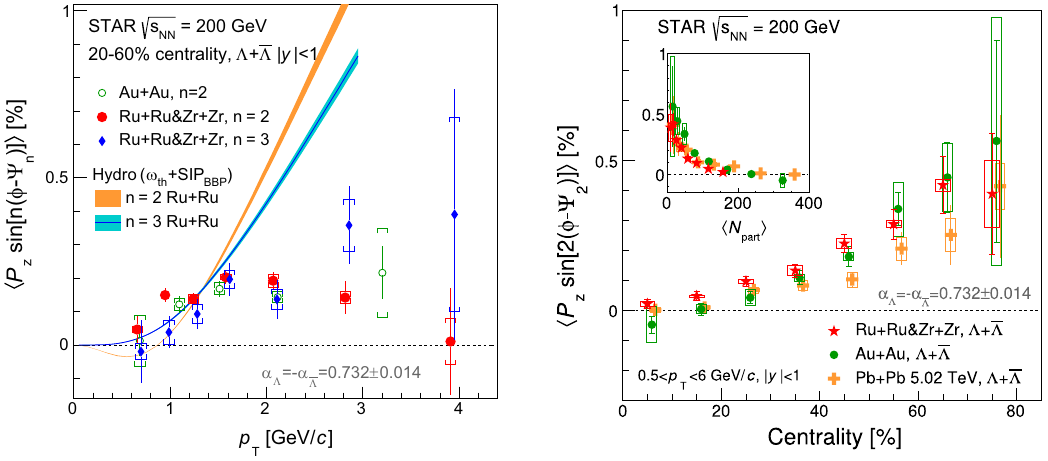}
\caption{$P_z$ sine coefficients of \lam+\alam hyperons as a function of transverse momentum $p_T$ (left) and centrality (right) in isobar collisions, comparing to hydrodynamic model calculations with one of the implementations for the shear-induced-polarization (SIP)~\cite{Alzhrani:2022dpi}. Also, the previous results in Au+Au collisions at 200 GeV and Pb+Pb collisions at 5.02 TeV are compared.
These figures are taken from Ref.~\cite{STAR:2023eck}.}
\label{fig:PzSin}
\end{center}
\end{figure}

The sine modulation was studied as a function of centrality and hyperons' transverse momentum $p_T$ as shown in Fig.~\ref{fig:PzSin}. The second- and third-order results are comparable and show a mild $p_T$ dependence. A slight difference at lower $p_T$ between the second and third orders looks similar to the relation between elliptic and triangular flow~\cite{PHENIX:2014uik}, which further supports the idea that the observed polarization is induced by anisotropic flow. Hydrodynamic model calculations with one of the existing implementations of the SIP~\cite{Alzhrani:2022dpi} lead to the correct sign and comparable magnitude to the data but not reproduce the detailed dependence, especially at higher $p_T$.
In the comparison with the results in Au+Au collisions at 200 GeV and Pb+Pb collisions at 5.02 TeV, there seems no significant collision system size nor energy dependence. It is worth mentioning that these measurements have sensitivity to specific shear and bulk viscosities as well as the initial condition~\cite{Alzhrani:2022dpi,Palermo:2022lvh}.

\section{Summary and Outlook}
Recent experimental results on the hyperon polarization in heavy-ion collisions have been reviewed. 
Although there have been a lot of progress in the polarization measurements, there still remain open questions to be understood. For the energy dependence of the global polarization, interestingly the global polarization seems to increase at the lower energy down to a few GeV. The behavior at the low energies, which may have sensitivity to the equation of state~\cite{Ivanov:2022ble}, should be explored with high precision. Global polarization of $\Xi$ hyperons was measured to be positive, though the significance is still $\sim2\sigma$ level. Global polarization of $\Omega$ as well as $\Xi$ hyperons will be explored in the coming RHIC runs.
Some of the differential measurements, e.g., rapidity and azimuthal angle dependence of polarization along the system angular momentum~\cite{Niida:2018hfw,BUR2023}, are predicted differently and need further investigation. New results on the polarization along the beam direction with the third-order event plane in isobar collisions confirm the picture of anisotropic-flow-driven polarization, which would provide constrains on the contributions from thermal vorticity and shear-induced polarization. Several interesting phenomena, e.g., spin Hall effect~\cite{Liu:2020dxg} and vortex ring~\cite{Voloshin:2017kqp,Lisa:2021zkj}, are predicted. The data of 200 GeV Au+Au collisions from 2023+2025 runs at RHIC as well as the future LHC runs and new experiments focusing on the high density region will produce high statistics data and be useful for studying the observables mentioned above.

\section*{Acknowledgements}
The author thanks S. Voloshin for fruitful discussion. This material is based upon the work supported by JSPS KAKENHI Grant Number JP22K03648.


\end{document}